\documentclass[mathleft]{an}
\usepackage{graphicx}
\usepackage{times}
\usepackage{natbib}
\begin{document}


\title{Peculiarities in the stellar velocity distribution of galaxies with a two-armed spiral structure}

\author{E. I. Vorobyov\inst{1,2}\fnmsep\thanks{Corresponding author:
  \email{vorobyov@ap.smu.ca}\newline}
\and  Ch. Theis\inst{3}
}
\titlerunning{Stellar velocity ellipsoids}
\authorrunning{E.I. Vorobyov \& Ch. Theis}
\institute{The Institute for Computational Astrophysics, Saint Mary's University, 
Halifax, NS, B3H 3C3, Canada 
\and 
Institute of Physics, South Federal University, Stachi 194, Rostov-on-Don, 344090, Russia
\and 
Institut f\"ur Astronomie, Universit\"at Wien,
              T\"urkenschanzstr. 17, 1180 Wien, Austria}

\received{}
\accepted{}
\publonline{}


\abstract{We expand our pervious numerical study of the properties of the stellar velocity distribution
within the disk of a two-armed spiral galaxy by considering spiral stellar density waves with 
different global Fourier amplitudes, $C_2$.
We confirm our previous conclusion that the ratio $\sigma_1:\sigma_2$ of smallest versus 
largest principal axes of the stellar velocity ellipsoid becomes abnormally small near the 
outer edges of the stellar spiral arms. The extent to which the stellar velocity ellipsoid is 
elongated (as compared to the unperturbed value typical for the axisymmetric disk) 
increases with the strength of the spiral density wave.
In particular, the $C_2=0.06$ spiral can decrease the unperturbed value of $\sigma_1:\sigma_2$ 
by $20\%$, while the $C_2=0.13$
spiral can decrease the unperturbed $\sigma_1:\sigma_2$ by a factor of 3. 
The abnormally small values of the $\sigma_1:\sigma_2$ ratio can potentially be used 
to track the position of {\it stellar} spiral density waves. The $\sigma_{\phi\phi}:\sigma_{rr}$
ratio is characterized by a more complex behaviour and exhibits less definite 
minima near the outer edges of the spiral arms.
We find that the epicycle approximation is violated near the spiral arms and 
cannot be used in spiral galaxies with $C_2\ga 0.05-0.06$ or in galaxies with the amplitude 
of the spiral stellar density wave (relative to the unperturbed background) of order
$0.1$ or greater.}

\maketitle

\section{Introduction}
It is well known that the local velocities of stars in galactic disks  are non-isotropic.
The velocity distribution of stars is usually described by the Schwarzschild distribution
function, which is a usual Gaussian distribution but with different squared velocity
dispersions $\sigma^2_{rr}$, $\sigma^2_{\phi\phi}$, and $\sigma^2_{zz}$ in the radial ($r$),
azimuthal ($\phi$), and vertical ($z$) coordinate directions. In addition, the cross-correlated
velocity dispersions (e.g. $\sigma^2_{r\phi}$ ) may be non-zero and the principal axes of
the velocity dispersion tensor $\sigma^2_{\rm ij}$ ($i,j=r,\phi,z$) 
may not be aligned with the coordinate axes.

In an axysimmetric, stationary disk stellar orbits are nearly circular
and characterized by epicycle motions with small radial amplitudes 
with respect to the guiding center. 
In such a disk the Oort ratio $X^2\equiv \sigma^2_{\phi\phi}/\sigma^2_{rr}$ 
can be expressed as
$X^2_{\rm ep}=-B/(A - B)$, where A and B are the usual Oort constants.
For the most interesting case of a flat rotation curve, this gives $X^2_{\rm ep}=0.5$.
However, the Solar Neighbourhood value for the Oort ratio $X^2_{\rm s}$ has been demonstrated by 
almost all observers to be less than $0.5$ \citep[see e.g.][]{Kerr86}. 
A possible explanation for this discrepancy between $X^2_{\rm ep}$ and $X^2_{\rm s}$ is 
the non-axysimmetric gravitational perturbation
from the Galactic bar that causes significant perturbations to the velocity moments 
\citep[e.g.][]{Muhlbauer}. In addition, the Oort constants  can be affected by the bar's potential
\citep[e.g.][]{Minchev07a}.

The gravitational field of stellar spiral arms can also gives rise to anomalies in the stellar 
velocity distribution \citep[e.g.][]{Mayor,Minchev07}. Recently, \citet{VT1}
have studied numerically the dynamics of stellar disks with a saturated two-armed spiral structure 
using the Boltzmann moment equation up to second order. They find that the spiral gravitational
field introduces large-scale non-circular motions in stellar orbits
near the outer edges of the spiral arms.
These stellar streams are responsible for the peculiar properties of the stellar velocity distribution
such as non-zero vertex deviations, abnormally low values of the ratio 
$\sigma_1:\sigma_2$ of smallest versus largest principal axes of the stellar velocity dispersion 
tensor, and large deviations of the ratio $\sigma^2_{\phi\phi}:\sigma^2_{rr}$ 
from those predicted by the epicycle approximation ($X^2_{\rm ep}$).

In this paper, we expand our previous numerical study of the stellar velocity distribution in
spiral galaxies with a two-armed structure \citep{VT1} by considering
stellar spiral density waves of different amplitude. 
The paper is organized as follows. The description of our model stellar disk is given in
Section~\ref{model}. The peculiar properties of the stellar velocity ellipsoid 
and the violation of the epicycle approximation are discussed in Sections~\ref{ellipsoid} and 
\ref{violation}, respectively. The main conclusions are summarized in Section~\ref{conclusions}.

\section{Model description}
\label{model}
We use the BEADS-2D code that is designed to study the dynamics of stellar disks
\citep{VT2}. The BEADS-2D code
is a finite-difference numerical code that solves the Boltzmann moment equations 
up to second order in the thin-disc approximation on the polar grid ($r,\phi$). 
More specifically, the BEADS-2D code solves for the stellar
surface density $\Sigma$, mean radial and azimuthal stellar velocities 
$u_{r}$ and $u_{\phi}$, and stellar velocity dispersion tensor $\sigma_{ij}$. The latter includes 
the squared radial and azimuthal stellar velocity dispersions $\sigma^2_{rr}$ and 
$\sigma^2_{\phi\phi}$, respectively,
and the mixed velocity dispersion $\sigma^2_{r\phi}$.
We close the system of Boltzmann moment equations by adopting the
zero-heat-flux approximation. 
For the main equations and necessary tests we refer the reader to \citet{VT2,VT1}.

Our model galaxy
consists of a thin, self-gravitating stellar disc embedded in
a static dark matter halo. The initial surface density of stars is axisymmetric and is
distributed exponentially according to 
\begin{equation}
\Sigma(r)=\Sigma_0 \exp(-r/r_{\rm d}),
\end{equation}
with a radial scale length $r_{\rm d}$ of 4~kpc and central surface density 
$\Sigma_0=10^3~M_\odot$~pc$^{-2}$. The gravitational potential of the stellar disk 
is calculated by solving for the Poisson integral using the convolution theorem 
\citep[][section 2.8]{BT}. 

The initial mean rotational (azimuthal) velocity of stars in the disc is chosen according to
\begin{equation}
   u_{\phi} = u_\infty \cdot \left( \frac{r}{r_{\rm flat}} \right)
                 \cdot 
                   {\displaystyle 
                    \left[ 1 + \left( \frac{r}{r_{\rm flat}} \right)^{n_v} 
                    \right]^{\displaystyle -\frac{1}{n_v}}} \,\,.
   \label{vcirc}
\end{equation}
The transition radius between an inner region of rigid rotation 
and a flat rotation in the outer part is given by $r_{\rm flat}$, which
we set to 3~kpc. The smoothness of the transition is controlled by the parameter
$n_v$, set to 3. The velocity at infinity ($u_\infty$) is set to 208~km~s$^{-1}$.

The radial component of the stellar velocity dispersion 
is obtained from the relation $\sigma_{rr}=3.36 \,Q_{\rm s}\, G \,\Sigma/\kappa$ 
for a given value of the Toomre parameter $Q_{\rm s}$. Here, $\kappa$ is the epicycle frequency. 
We assume that throughout 
most of the disc $Q_{\rm s}$ is constant and equal 1.3 but is steeply increasing 
with radius at $r>25$~kpc. The azimuthal 
component of the velocity dispersion $\sigma_{\phi\phi}$ is determined from
$X^2_{\rm ep}$ using the initial rotation velocity of stars $u_{\phi}$
in the Oort constants. Once the rotation curve and the radial 
profiles of the stellar surface density and velocity
dispersions are fixed, the dark matter halo potential  
can be derived from the steady-state momentum equation for the azimuthal velocity of stars
\citep[see][for details]{VT1}.

In the current simulations, the numerical resolution has $512\times 512$ 
grid zones that are equally spaced in the $\phi$-direction and logarithmically spaced 
in the $r$-direction. The inner and outer reflecting boundaries are at 
$r_{\rm in}=0.2$~kpc and $r_{\rm out}=35$~kpc, 
respectively. The total disk mass $M_{\rm st}=10^{11}~M_\odot$.
We extend the outer computational boundary far enough to ensure a good 
radial resolution in the region of interest within the inner 20~kpc. For instance, 
the radial resolution at 1~kpc and 20~kpc is approximately 10~pc and 150~pc,
respectively.

\section{Peculiar shape of the stellar velocity ellipsoid}
\label{ellipsoid}
\begin{figure}
   \resizebox{\hsize}{!}{
     \includegraphics[angle=0]{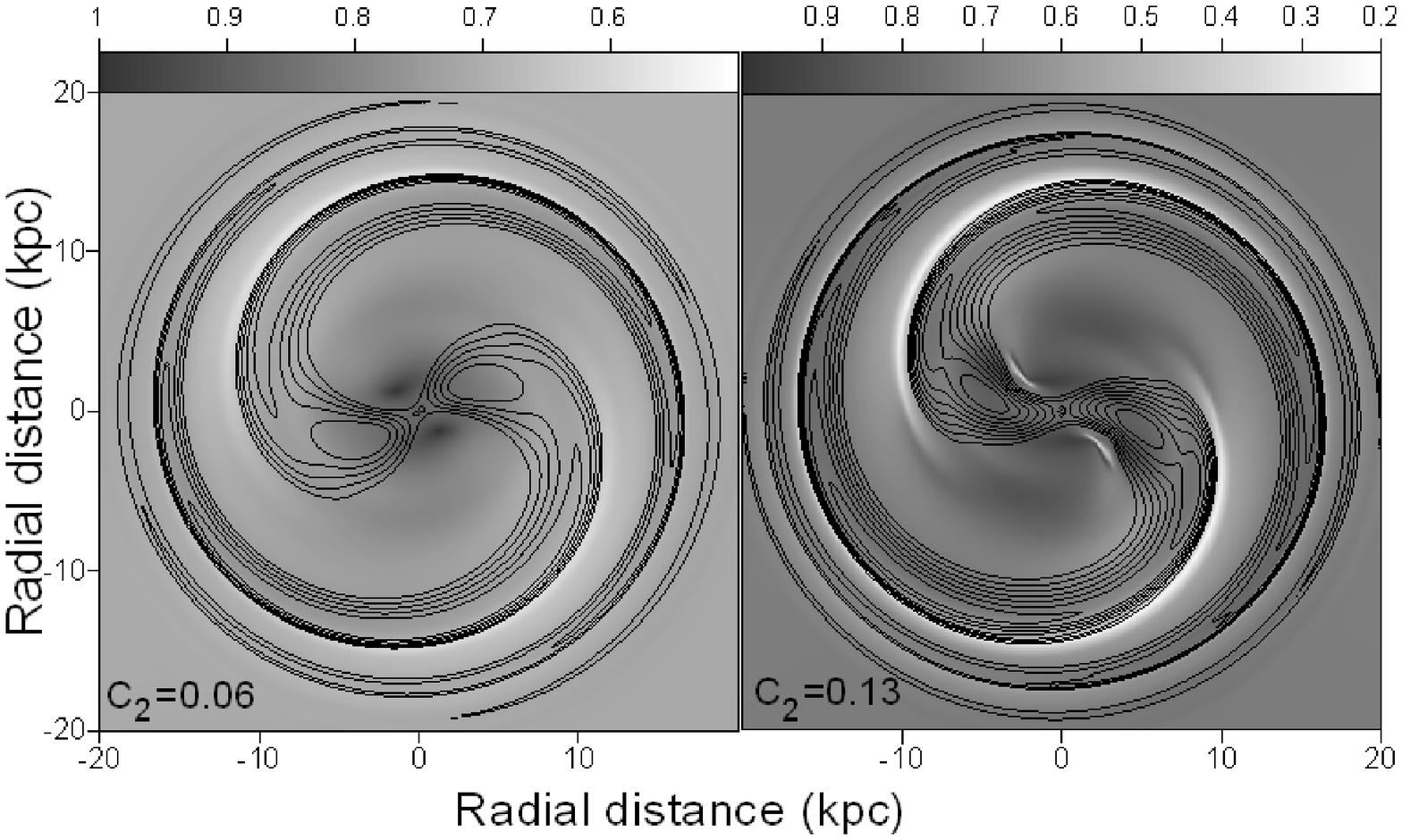}}     
   \caption{Relative stellar density perturbation (black contour lines) superimposed 
on the grey-scale image of the $\sigma_1:\sigma_2$ distribution at $t=1.5$~Gyr (left) and 
$t=1.6$~Gyr (right).
The corresponding global Fourier amplitudes of the m=2 mode ($C_2$) are indicated in each panel.}
   \label{fig1}
\end{figure}
In the beginning of our simulations we add a small ($\le 10^{-5}$) random perturbation to 
the initially axisymmetric surface density distribution of stars. 
Since the model stellar disc is characterized by $Q_{\rm s}=1.3$, 
it becomes vigorously unstable to the growth of non-axisymmetric gravitational instabilities.
The initial parameters of our model are chosen to favour the growth of a two-armed 
spiral pattern. The most likely 
physical interpretation for the growth of a spiral structure in our model disk
is swing amplification \citep{VT2,VT1}. 
The black contour lines in Fig.~\ref{fig1} show the relative stellar density perturbation
defined as
\begin{equation}
\zeta(r,\phi) = {\Sigma(r,\phi) - \Sigma_{\rm in}(r) \over \Sigma_{\rm
in}(r) },
\end{equation}
where $\Sigma_{\rm in}(r)$ is the initial axisymmetric stellar density
distribution in the disk and $\Sigma(r,\phi)$ is the current stellar density
distribution. Only positive density perturbations are plotted. The left panel in Fig.~\ref{fig1}
corresponds to $t=1.5$ Gyr since the beginning of our simulations and the minimum and maximum contour
levels correspond to the relative perturbations of 0.03 and 0.23, respectively. 
The right panel shows the relative stellar density perturbations at $t=1.6$~Gyr.
The minimum and maximum contour levels are $\zeta=0.05$ and $\zeta=0.75$, respectively.
To quantify the strength of the spiral arms, we use the global Fourier amplitudes defined as
\begin{equation}
C_{\rm m} (t) = {1 \over M_{\rm d}} \left| \int_0^{2 \pi} \int_{r_{\rm in}}^{r_{\rm out}} 
\Sigma(r,\phi,t) \, e^{im\phi} r \, dr\,  d\phi \right|.
\end{equation}
The dominant $m=2$ mode in the left/right panels of Fig.~\ref{fig1} is characterized by the
$C_2=0.06$ and $C_2=0.13$, respectively.

The velocity dispersion tensor $\sigma^2_{ij}$
is often non-diagonal in the local coordinate system ($r,\phi, z$).
The principal axes of a {\it diagonalized} velocity dispersion tensor 
form an imaginary ellipsoidal surface that is called the velocity ellipsoid.
The available measurements in the solar vicinity
indicate that the ratio $\sigma_1:\sigma_2$ of smallest versus largest principal 
axes of stellar velocity ellipsoids in the disk plane does not vary significantly 
with the $B-V$ colour or with the age stellar populations \citep{Dehnen}. 
On the other hand, the $\sigma_1:\sigma_2$ ratio is expected to have deep minima
near the outer edges of spiral arms and this peculiar, elongated shape of 
the stellar velocity ellipsoid can be used to track the position
of spiral stellar density waves \citep{VT1}.  In the present paper we corroborate our previous
results by considering spiral stellar density waves with different 
global Fourier amplitudes. The grey-scale images in Fig.~\ref{fig1} show the spatial distribution 
of $\sigma_1:\sigma_2$ obtained at $t=1.5$~Gyr (left), when the global Fourier amplitude 
of the $m=2$ mode is $C_2=0.06$, and at $t=1.6$~Gyr (right), 
when $C_2$ has reached a maximum value of $0.13$. It is seen that the stronger spiral 
produces a larger response in the shape of the stellar velocity ellipsoids.
For instance, the minimum values of $\sigma_1:\sigma_2$ in the  $C_2=0.06$ disk 
are about $0.5-0.6$, but they decrease to $0.2-0.3$ in the $C_2=0.13$ disk.
In both cases the shape of the stellar velocity ellipsoids is most elongated near the outer edges of
the spiral arms, with the exception of a small region near the central bar.

\begin{figure}
   \resizebox{\hsize}{!}{
     \includegraphics[angle=0]{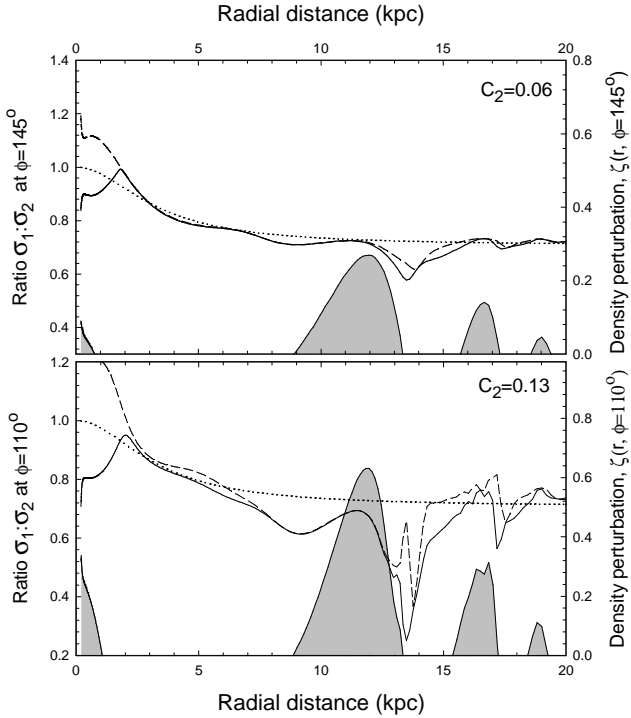}}     
   \caption{Radial profiles of the $\sigma_1:\sigma_2$ ratio (solid lines), 
   $\sigma_{\phi\phi}:\sigma_{rr}$ 
    ratio (dashed lines), and relative stellar density perturbation
    $\zeta$ (shaded area) obtained by taking a narrow radial cut at $\phi=145^\circ$ 
in the top panel of Fig.~\ref{fig1} (top) and $\phi=110^\circ$ in the bottom panel 
of Fig.~\ref{fig1} (bottom). The dotted lines present the unperturbed initial values of 
$\sigma_1:\sigma_2$. The corresponding strength of the $m=2$ mode is indicated in each panel.}
   \label{fig2}
\end{figure}

To better illustrate this phenomenon, we take a narrow radial cut in Fig.~\ref{fig1}
at $\phi=145^\circ$ (top) and $\phi=110^\circ$ (bottom), and plot the resulting radial distributions
of $\zeta$ (shaded area) and $\sigma_1:\sigma_2$ (solid lines) in Fig.~\ref{fig2}. 
The angle $\phi$ is counted counterclockwise from the positive horizontal axis of Fig.~\ref{fig1}.
The dotted line shows the initial unperturbed values of $\sigma_1:\sigma_2$. It is evident
that the weaker $C_2=0.06$ spiral (top) causes only a mild perturbation to 
the initial shape of the stellar velocity ellipsoids. However, there is a noticeable
decrease in the values of $\sigma_1:\sigma_2$ near the outer edge of the innermost (and strongest)
spiral arm, indicating that the shape of the stellar velocity ellipsoid becomes abnormally elongated there.
As the global Fourier amplitude of the $m=2$ mode increases to $C_2=0.13$ (bottom),
a considerable change in the initial shape of the stellar velocity ellipsoid becomes evident.
Deep minima in the radial distribution of $\sigma_1:\sigma_2$ have occurred near the outer 
edges of the spiral arms. We conclude that the peculiar elongated shape of the stellar velocity ellipsoids
can be used as a tracer of the stellar spiral density wave if the latter is sufficiently strong, 
$C_2 \ga 0.1$.


It may be tempting to use the ratio
$\sigma_{\phi\phi}:\sigma_{rr}$ instead of $\sigma_1:\sigma_2$ in order to track the position of stellar
spiral density waves. However, we find that the latter ratio may not be as good a diagnostic tool 
as the former ratio.
Indeed, the dashed lines in Fig.~\ref{fig2} show the radial distribution of the
$\sigma_{\phi\phi}:\sigma_{rr}$ ratio taken along the same azimuthal angles as 
in the case of the $\sigma_1:\sigma_2$ ratio. It is evident that the 
$\sigma_{\phi\phi}:\sigma_{rr}$ ratio also exhibits local minima near the
outer edges of the spiral arms but these depressions may not be as deep as those
of the $\sigma_1:\sigma_2$ ratio.

\section{Violation of the epicycle approximation}
\label{violation}

\begin{figure}
   \resizebox{\hsize}{!}{
     \includegraphics[angle=0]{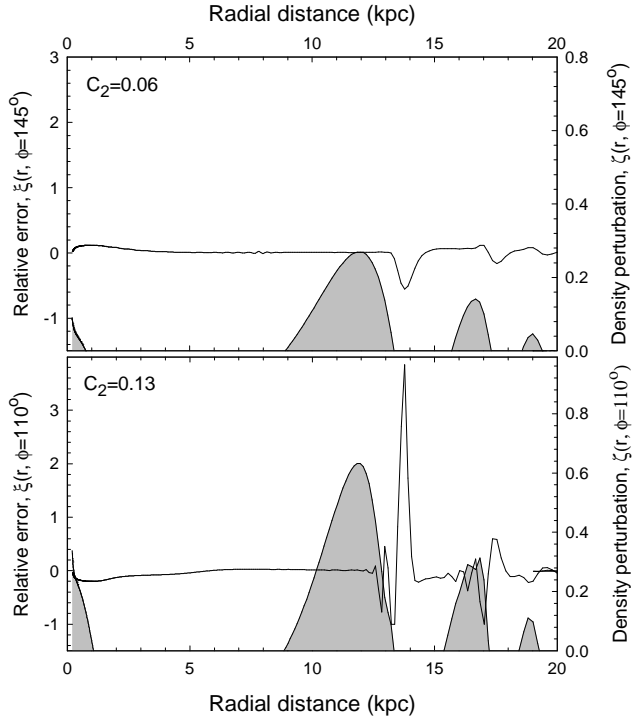}}     
   \caption{Radial profiles of $\xi$ (solid lines) and relative stellar density perturbation
    $\zeta$ (shaded area) obtained by taking a narrow radial cut at $\phi=145^\circ$ 
    in the top panel of Fig.~\ref{fig1} (top) and $\phi=110^\circ$ in the bottom panel of 
    Fig.~\ref{fig1} (bottom). The corresponding strength of the $m=2$ mode is indicated in each panel.}
   \label{fig3}
\end{figure}

It is often observationally difficult to measure both radial and azimuthal velocity dispersions of stars. 
Hence, it is tempting to use the Oort ratio to derive either of the two dispersions from the known
one and the circular speed of stars ($u_{\rm c}$)
\begin{equation}
X^2= {-B \over A-B}= {1 \over 2} \left( {r\over u_{\rm c}} {du_{\rm c} \over dr} +1 \right).
\label{epicycle}
\end{equation}
One has to keep in mind that
equation~(\ref{epicycle}) may be valid only in the epicycle approximation
\citep[e.g][]{Kuijken}

To test the validity of equation~(\ref{epicycle}), we calculate the following quantity that
measured the relative deviation from the epicycle approximation
\begin{equation}
\xi(r,\phi)= {X^2 - \sigma_{\phi\phi}:\sigma_{rr} \over \sigma_{\phi\phi}:\sigma_{rr}},
\end{equation}
where the ratio $\sigma_{\phi\phi}:\sigma_{\rm rr}$ is calculated 
from the model's known velocity dispersions and $X^2$ is calculated using equation~(\ref{epicycle})
and $u_{\phi}$ as a proxy for $u_{\rm c}$.
The resulting radial profiles of
$\xi$ taken along the azimuthal angles $\phi=145^\circ$ (top) and $\phi=110^\circ$ (bottom)
are plotted in Fig.~\ref{fig3} by the solid lines. The shaded area shows the positive stellar density
perturbation $\zeta$ for the $C_2=0.06$ spiral (top) and $C_2=0.13$ spiral (bottom), respectively.
It is seen that the epicycle approximation is violated near the outer edges of the spiral arms.
The extent to which the epicycle approximation is violated depends on the strength 
of the spiral structure. For instance, the $C_2=0.06$ spiral shows a moderate deviation of
order $50\%$ near the position of the innermost spiral arm and a mild deviation of order $20\%$
near the second spiral arm. As the strength of the dominant $m=2$ mode has increased to
$C_2=0.13$, the deviation from the epicycle approximation begins to exceed $100\%$.
{\it It is therefore very risky to use the epicycle approximation to derive velocity dispersions,
when the global Fourier amplitude of the dominant spiral mode exceeds $0.05-0.06$.}
Figure~\ref{fig3} indicates that the same criterion can be expressed in terms of $\zeta$ -- the 
relative amplitude of the spiral stellar density wave should not exceed $0.1$.

\section{Conclusions}
\label{conclusions}
We have studied numerically the stellar velocity distribution in a two-armed spiral galaxy
with a varying strength of the dominant $m=2$ mode. The development of the spiral 
structure and its time evolution are followed self-consistently by solving the 
Boltzmann moment equations up to second order in the thin-disk approximation. We find the following:
\begin{itemize}
\item The stellar spiral density wave produces peculiar signatures in the velocity distribution
of the underlying stellar disk. For instance, the ratio $\sigma_1:\sigma_2$ of the smallest versus 
largest principal axes of the stellar velocity ellipsoid becomes abnormally small (as compared to 
the unperturbed value typical for an axysimmetric 
stellar disk) near the outer edges of the spiral arms. The degree to which the stellar velocity 
ellipsoid becomes elongated depends on the strength of the spiral arms (as defined by the global 
Fourier amplitudes, $C_{\rm m}$). 
For instance, the $C_2=0.06$ spiral can reduce the initial (unperturbed) value of 
$\sigma_1:\sigma_2$ from 0.75 to 0.6, while the $C_2=0.13$ spiral can reduce the initial 
$\sigma_1:\sigma_2$ by a factor of three.
The abnormally small values of the $\sigma_1:\sigma_2$ ratio can potentially be used 
to track the position of {\it stellar} spiral density waves.
\item The epicycle approximation is violated near the spiral arms of galaxies with the global 
Fourier amplitude of the dominant $m=2$ mode of order $C_2=0.05-0.06$ or greater. This corresponds 
to the stellar density perturbation (relative to the unperturbed background disk) of order 0.1 or 
greater. 
\item The $\sigma_{\phi\phi}:\sigma_{rr}$ ratio of the azimuthal to 
radial velocity dispersions shows less definite minima near the outer edges of the spiral arms.

\end{itemize}

\acknowledgements
E.I.V. gratefully acknowledges support from an ACEnet Fellowship.
The numerical simulations were performed on the Atlantic Computational Excellence 
Network (ACEnet). C.T. is grateful to the priority program {\it Rechnergest\"utzte
Wissenschaften} of the Univ.\ of Vienna for financial support.


\end{document}